\documentclass[aps,showpacs,amsmath,amssymb]{revtex4}

\usepackage[dvips]{graphicx}

\newcommand{\bra}[1]{\mbox{$\left\langle{#1}\right|$}}
\newcommand{\ket}[1]{\mbox{$\left|{#1}\right\rangle$}}

\begin{document}

\title{Combined local implementation of nonlocal operations using GHZ states}
\author{Ning Bo Zhao\footnote{nbzhao@mail.ustc.edu.cn}, An Min Wang\footnote{anmwang@ustc.edu.cn}}
\affiliation{Quantum Theory Group, Department of Modern Physics\\
University of Science and Technology of China, Hefei 230026, People
Republic of China}

\begin{abstract}
We propose a protocol for local implementation of two consecutive nonlocal operations by three parters.
It consumes one shared GHZ state in this protocol.
We also demonstrate that these resources are sufficient and necessary to locally implement two consecutive CNOT operations.
\end{abstract}


\maketitle
\section{Introduction}
In distributed quantum computation, collective operations often need to be implemented on the qubits at distant nodes.
Such operations can only be implemented locally, i.e., it can only be implemented using local operations and classical communications (LOCC), shared entanglement and some auxiliary qubits.
If the circuit of the operation has been constructed in one node, a straightforward method to implement such a nonlocal operation is using quantum state teleportations \cite{six93}, i.e., teleporting all of the qubits to this node, performing the operation and teleporting these qubits back.

In the bipartite case, two rounds of state teleportation are required using this method.
So in this case, it consumes two ebits (shared entanglement resources) .
However, there are operations that can be locally implemented using less entanglement resources \cite{p00,clp01}.
It has been demonstrated theoretically \cite{p00} and experimentally \cite{guo04} that in order to implement a CNOT operation, one ebit is necessary and sufficient.
In fact, any diagonal or offdiagonal block operation can be implemented using the same resources, even if the detail of the blocks are unknown \cite{my08}.
Recently, the problems of constructing a nonlocal operation or simulating it by other operations are discussed \cite{d1,d2,d3,rag,guo05}.
In this paper, we only consider the case that the circuit of the operation has been locally constructed in one node.
Of course, if the operation is completely known, it can also be implemented using the simulating methods discussed the in the above references.
But, the parters must construct new circuits to use these methods.
These accessorial devices need consume more local resources and may bring a loss of accuracy.

In the multiparty case, effective protocols for some operations are also proposed recently \cite{p00,my08}.
The entanglement resources used in these protocols are Bell states.
Nevertheless, in multiparty case, the GHZ states are also important and useful.
So it is also interesting to search effective protocols using GHZ states.

There are some similarities between the local implementation of nonlocal operations and the remote implementation of local operations.
Usually, a protocol for remote implementation of operations with some forms can also be used to locally implement nonlocal block operations with similar forms \cite{my08}.
Ref. \cite{wang07} proposed a protocol for consecutively remote implementation of two diagonal operations from two distant nodes to one node, using a shared GHZ state.
This method can also be used in local implementation of nonlocal operations.
We will discuss this issue in the following sections.

In Sec. \ref{p3}, we propose a protocol to locally implement two consecutive diagonal block operations among three parters --- the first is on Alice's qubit $A$ and Bob's qubit $B$, the second is on $A$ and Charlie's qubit $C$.
It requires one shared GHZ state, and it is available even if the detail of the blocks are unknown.
The CONT operation is a diagonal block operation, so two consecutive CNOTs can also be implemented using the protocol.
We prove that these resources (one shared GHZ state) are also necessary for consecutive two CNOT operations.
In Sec. \ref{con}, we summarize our results.
\section{the protocol} \label{p3}
Three distant parters Alice, Bob and Charlie have three qubits $A$, $B$ and $C$ respectively.
Alice and Bob need to implement a CNOT on qubits $A$ and $B$.
Following the protocol in Ref. \cite{p00}, Alice and Bob can accomplish this task using one shared Bell state (one ebit) --- the necessary and sufficient entanglement resources for this task.
Then consecutively, Alice and Charlie need to implement a CNOT on qubits $A$ and $C$.
Similarly, Alice and Charlie need another shared Bell state to accomplish this task.
Thus, if considering these two operations as a whole, using above method, the three parters need a total of two shared Bell state --- one shared by Alice and Bob, the other shared by Alice and Charlie.
However, they can also implement these two CNOT operations using only one GHZ state shared by the three parters.
We will propose such a protocol in this section.
This protocol can be used to consecutively implement any two diagonal block operations even if the detail of the blocks are unknown.
Since the CNOT operation is also a diagonal block operation, we will discuss the general case firstly.
Then we will discuss the necessary entanglement resources for two consecutive CNOTs.

Alice, Bob and Charlie need to implement two nonlocal operations in turn.
Firstly, Alice and Bob need to implement a two-qubit diagonal block operation $U$ on their qubits $A$ and $B$, where
\begin{equation}\label{u}
U =
\left( \begin{array}{cc}
u_0 & 0\\
0 & u_1
\end{array} \right)=\sum_{i=0}^{1} \ket{i}\bra{i} \otimes u_i,
\end{equation}
where $u_0$ and $u_1$ are $2\times 2$ unitary matrices.
Bob has the device of the operation $U$, but he may only know the form of it (Eq. \ref{u}), and may not know the detail of $u_0$ and $u_1$.
Then, Alice and Charlie need to implement a two-qubit diagonal block operation $V$ on their qubits $A$ and $C$, where
\begin{equation}\label{v}
V =
\left( \begin{array}{cc}
v_0 & 0\\
0 & v_1
\end{array} \right)=\sum_{i=0}^{1} \ket{i}\bra{i} \otimes v_i,
\end{equation}
where $v_0$ and $v_1$ are $2\times 2$ unitary matrices.
Charlie has the device of the operation $V$, but he may only know the form of it (Eq. \ref{v}), and may not know the detail of $v_0$ and $v_1$.
Thus, as a result, they need to implement a three-qubit operation
\begin{eqnarray}
W &=& \left(\sum_{i=0}^{1} \ket{i}\bra{i} \otimes I \otimes v_i \right) \left(\sum_{j=0}^{1} \ket{j}\bra{j} \otimes u_j \otimes I\right) \nonumber \\
&=& \sum_{i=0}^{1} \ket{i}\bra{i} \otimes u_i \otimes v_i
\end{eqnarray}
on their qubits $A,B,C$.

These two nonlocal operation can be locally implemented in turn using the protocol proposed in Ref. \cite{my08}.
Using this method, they need two Bell states --- one shared by Alice and Bob, the other shared by Alice and Charlie.
We will prove in the followings that they can also locally implement these two operations using one GHZ state shared by the three parters.

\paragraph*{Initial state}
The initial state of the qubits $A,B,C$ can always be expressed as
\begin{equation}
\ket{\Psi_0}_{ABC}=\alpha_0 \ket{0}_A \ket{\xi_0}_{BC} +\alpha_1 \ket{1}_A \ket{\xi_1}_{BC}, 
\end{equation}
where $\ket{\xi_0}$ and $\ket{\xi_1}$ can be any two-qubit state and are not orthogonal in general.

They share a GHZ state in qubits $A_1B_1C_1$ 
\begin{equation}\label{e}
\ket{\Phi}_{A_1B_1C_1}=\frac{1}{\sqrt{2}}(\ket{000}+\ket{111}),
\end{equation}
where qubit $A_1$ belongs to Alice, qubit $B_1$ belongs to Bob and qubit $C_1$ belongs to Charlie.

They can use the protocol specified as the following steps to accomplish their tasks.

\paragraph*{step 1}
Alice performs a CNOT operation on her qubits $A$ and $A_1$, using the qubit $A$ as the control.
After this, the state of $A,B,C,A_1,B_1,C_1$ becomes
\begin{eqnarray}
& & CNOT^{A,A_1} \ket{\Psi_0}_{ABC} \otimes \ket{\Phi}_{A_1B_1C_1} \nonumber \\
&=& CNOT^{A,A_1} \sum_i \alpha_i \ket{i}_A \ket{\xi_i}_{BC} \frac{1}{\sqrt{2}} \sum_j \ket{jjj}_{A_1B_1C_1}  \nonumber \\
&=& \frac{1}{\sqrt{2}} \sum_{ij} \alpha_i \ket{i}_A \ket{\xi_i}_{BC} \ket{i\oplus j}_{A_1} \ket{jj}_{B_1C_1},
\end{eqnarray}
where ``$\oplus$'' denotes the addition module 2.

Then she measures $A_1$ in computational basis $\ket{a}\bra{a},(a=0,1)$, and tell the result $a$ to Bob and Charlie.
The state of $A,B,C,B_1,C_1$ becomes
\begin{eqnarray}
& & \sum_{ij} \alpha_i \ket{i}_A \ket{\xi_i}_{BC} \ket{jj}_{B_1C_1} \delta_{a,i\oplus j}\nonumber \\
&=& \sum_{ij} \alpha_i \ket{i}_A \ket{\xi_i}_{BC} \ket{jj}_{B_1C_1} \delta_{j,i\oplus a} \nonumber \\
&=& \sum_{i} \alpha_i \ket{i}_A \ket{\xi_i}_B \ket{i\oplus a}_{B_1} \ket{i\oplus a}_{C_1}.
\end{eqnarray}

\paragraph*{step 2}
If the result $a=0$ Bob does nothing, if $a=1$ Bob performs the operation $X$ on $B_1$, where
\begin{equation}
X =
\left( \begin{array}{cc}
0 & 1\\
1 & 0
\end{array} \right)
\end{equation}
is the first Pauli matrix.
Because $X\ket{i}=\ket{i\oplus 1},(i=0,1)$, the state of $A,B,C,B_1,C_1$ becomes
\begin{equation}
\sum_{i} \alpha_i \ket{i}_A \ket{\xi_i}_{BC} \ket{i}_{B_1} \ket{i\oplus a}_{C_1}.
\end{equation}

\paragraph*{step 3}
Bob performs the two-qubit operation $U$ on his qubits $B_1$ and $B$.
The state of $A,B,C,B_1,C_1$ becomes
\begin{eqnarray}
& & U^{B_1,B} \sum_{i} \alpha_i \ket{i}_A \ket{\xi_i}_{BC} \ket{i}_{B_1} \ket{i\oplus a}_{C_1} \nonumber \\
&=& \sum_{i} \alpha_i \ket{i}_A [(u_i\otimes I)\ket{\xi_i}]_{BC} \ket{i}_{B_1} \ket{i\oplus a}_{C_1}.
\end{eqnarray}

\paragraph*{step 4}
Bob performs the operation $H$ on $B_1$, where
\begin{equation}
H =
\left( \begin{array}{cc}
1 & 1\\
1 & -1
\end{array} \right)
\end{equation}
is the Hadamard operation.
Because $H\ket{i}=\frac{1}{\sqrt{2}}(\ket{0}+(-1)^i\ket{1})$, the state of $A,B,C,B_1,C_1$ becomes
\begin{eqnarray}
& & \sum_{i} \alpha_i \ket{i}_A \left[(u_i\otimes I)\ket{\xi_i}\right]_{BC} \frac{1}{\sqrt{2}}(\ket{0}+(-1)^i\ket{1})_{B_1} \ket{i\oplus a}_{C_1} \nonumber \\
&=& \frac{1}{\sqrt{2}} \ket{0}_{B_1} \sum_{i} \alpha_i \ket{i}_A \left[ (u_i\otimes I)\ket{\xi_i}\right]_{BC} \ket{i\oplus a}_{C_1}  \nonumber \\
& & + \frac{1}{\sqrt{2}} \ket{1}_{B_1} \sum_{i} (-1)^i \alpha_i \ket{i}_A \left[ (u_i\otimes I)\ket{\xi_i}\right]_{BC} \ket{i\oplus a}_{C_1} .
\end{eqnarray}

Bob measures $B_1$ in the computational basis $\ket{b}\bra{b}, (b=0,1)$, and tell the result $b$ to Alice.
The state of $A,B,C,C_1$ becomes
\begin{equation}
\sum_{i} (-1)^{ib} \alpha_i \ket{i}_A [(u_i\otimes I)\ket{\xi_i}]_{BC} \ket{i\oplus a}_{C_1}.
\end{equation}

\paragraph*{step 2'}
Charlie does the same thing on $C_1$ as Bob does on $B_1$ in step 2.

\paragraph*{step 3'}
Charlie performs the two-qubit operation $V$ on his qubits $C_1$ and $C$.

\paragraph*{step 4'}
Charlie does the same thing on $C_1$ as Bob does on $B_1$ in step 4.
Then he tells Alice the measurement result $c$.
Obviously, Charlie's operations in step 2'-4' and Bob's operations in step 2-4 are not correlative.
So, they can do these operations in parallel.

Similarly, after step 2'-4', the state of $A,B,C$ becomes
\begin{equation}
\sum_{i} (-1)^{i(b+c)} \alpha_i \ket{i}_A [(u_i\otimes v_i)\ket{\xi_i}]_{BC}.
\end{equation}

\paragraph*{step 5}
If $b=c$ $(b\oplus c=0)$ Alice does nothing, if $b\neq c$ $(b\oplus c=1)$ Alice performs the operation $Z$ on $A$, where
\begin{equation}
Z =
\left( \begin{array}{cc}
1 & 0\\
0 & -1
\end{array} \right)
\end{equation}
is the third Pauli matrix.
Because $Z\ket{i}=(-1)^i\ket{i}$, the state of $A,B$ becomes
\begin{equation}
\sum_{i} \alpha_i \ket{i}_A [(u_i\otimes v_i)\ket{\xi_i}]_{BC}.
\end{equation}
It can be directly calculated that
\begin{equation}
W \ket{\Psi_0}_{ABC} = \sum_{i} \alpha_i \ket{i}_A [(u_i\otimes v_i)\ket{\xi_i}]_{BC}.
\end{equation}

Thus, after these steps, these two diagonal block operation $U^{AB}$ and $V^{AC}$ are determinately implemented on $A,B,C$ using one shared GHZ state plus 3 cbits --- one from Alice to Bob and Charlie, one from Bob to Alice and one from Charlie to Alice.

\begin{figure}[ht]
\begin{center}
\includegraphics[scale=1.0]{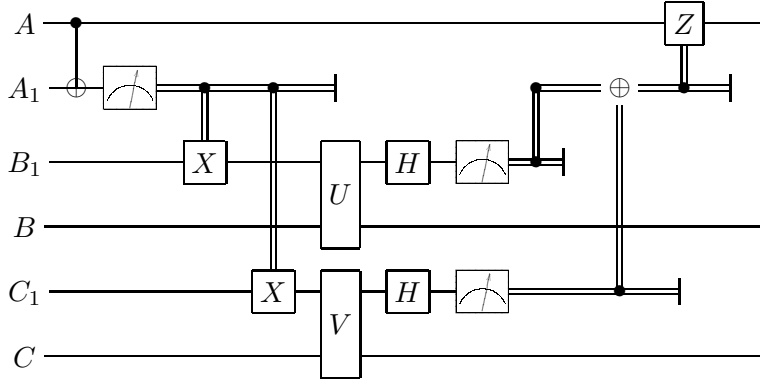}
\end{center}
\vskip -0.1in \caption{ \label{fig1}
Quantum circuit of the protocol using GHZ state, 
$A_1B_1C_1$ are in GHZ state defined by Eq. \ref{e}.}
\end{figure}

This protocol can be expressed as Fig. \ref{fig1}.

It can be found from the proof of the protocol that the protocol is independent on the dimension of the blocks.
So this protocol is also available if the blocks have higher dimensions, i.e., $u_0$ and $u_1$ are $2^N\times 2^N$ unitary matrices, $v_0$ and $v_1$ are $2^M\times 2^M$ unitary matrices, and Bob and Charlie have the corresponding $N$ and $M$ qubits.

Obviously, the CNOT operation is a diagonal block operation.
So two consecutive CNOT operations can be implemented using this method.
In fact, it equals that the three parters implement a three-qubit operation 
\begin{equation}
G=\ket{0}\bra{0} \otimes I \otimes I + \ket{1}\bra{1} \otimes X \otimes X 
\end{equation}
on their qubits $A,B,C$.

Suppose the three parters can implement $G$ on $ABC$ via some procedure using LOCC and shared entanglement resources.
Because
\begin{equation}
G (\ket{+}_A\ket{0}_B\ket{0}_C) = \frac{1}{\sqrt{2}}(\ket{000}+\ket{111})_{ABC}, \quad (\ket{+}=\frac{1}{\sqrt{2}}(\ket{0}+\ket{1}),
\end{equation}
so if the initial state of qubits $ABC$ is $\ket{+}_A\ket{0}_B\ket{0}_C$, the three parters can finally get a GHZ state in qubits $ABC$ after this procedure.
Thus, they require to share entanglement resources more than one GHZ state in the beginning, as the amount of entanglement can not be increased by LOCCs.

Thus, our protocol is optimal for two consecutive CNOT operations.
But, if they implement two CNOT consecutively using the protocol in Ref. \cite{p00}, they require two shared Bell states.
Obviously, they can get a shared GHZ state using these two Bell states via LOCC.
If regarding Bob and Charlie as a whole, the amount of bipartite entanglement between Alice and them is one ebit in one GHZ state, and is two ebits in two Bell states.
So, in this sense they need more entanglement resources than one shared GHZ state.
Thus it is not optimal to use this method.

\section{conclusions} \label{con}
In this paper, we have addressed the problem of the local implementation of nonlocal operations by three parters --- Alice, Bob and Charlie.
We have proposed a protocol for local implementation of two consecutive two-qubit diagonal block operations --- the first by Alice and Bob, the second by Alice and Charlie.
It consumes one shared GHZ state in this protocol.
However, they need two shared Bell state if they independently implement these two operations via the bipartite protocol in turn.
It is interesting that such overall treatment can save resources.
We have also demonstrated that one shared GHZ state is sufficient and necessary to locally implement two consecutive CNOT.
\section*{Acknowledgments}

We acknowledge all the collaborators of our quantum theory group at the Institute for Theoretical Physics of our university. 
This work was funded by the National Natural Science Foundation of China under Grant No. 60573008.


\end{document}